\documentclass[prb,aps,amssymb,showpacs,twocolumn]{revtex4}
\usepackage{amsmath}
\usepackage{amssymb}
\usepackage{amsthm}
\usepackage{amsfonts}
\usepackage{enumerate}
\usepackage{latexsym}
\usepackage{graphicx}

\newcommand{\beq}{\begin{equation}}
\newcommand{\eneq}{\end{equation}}

\input{epsf}

\begin{document}

\tolerance 10000

\newcommand{\vk}{{\bf k}}


\title{Diffusive-Ballistic Crossover and the Persistent Spin Helix}
\author{B. Andrei Bernevig  }
\affiliation{Princeton Center for Theoretical Physics, Princeton
University, Princeton, NJ 08544} \affiliation{Department of Physics,
Jadwin Hall, Princeton University, Princeton, NJ 08544}
\author{ Jiangping
Hu} \affiliation{Department of Physics, Purdue University, West
Lafayette, IN 47907}

\begin{abstract}
Conventional transport theory focuses on either the diffusive or
ballistic regimes and neglects the crossover region between the two.
In the presence of spin-orbit coupling, the transport equations are
known only in the diffusive regime, where the spin precession angle
is small. In this paper, we develop a semiclassical theory of
transport valid throughout the diffusive - ballistic crossover of a
special $SU(2)$ symmetric spin-orbit coupled system. The theory is
also valid in the physically interesting regime where the spin
precession angle is large. We obtain exact expressions for the
density and spin structure factors in both $2$ and $3$ dimensional
samples with spin-orbit coupling.
\end{abstract}

\date{\today}


\maketitle

The physics of systems with spin-orbit coupling has generated great
interest from both academic and practical perspectives
[\onlinecite{wolf2001}]. Spin-orbit coupling allows for purely
electric manipulation of the electron spin
[\onlinecite{nitta1997,grundler2000,kato2004A,kato2004,weber2005}],
and could be of practical use in areas from spintronics to quantum
computing. Theoretically, spin-orbit coupling is essential to the
proposal of interesting effects and new phases of matter such as the
intrinsic and quantum spin Hall effect
[\onlinecite{murakami2003,sinova2004,kane2005,kane2005A,bernevig2006,
bernevig2006science}[.

While the diffusive transport theory for a system with spin-orbit
coupling has recently been derived
[\onlinecite{burkov2003,mishchenko2004}], the analysis of
diffusive-ballistic transport - where the spin precession angle
during a mean free path is comparable to (or larger than) $2 \pi$ -
has so far remained confined to numerical methods
[\onlinecite{nomura2005}]. This situation is experimentally relevant
since the momentum relaxation time $\tau$ in high-mobility GaAs or
other semiconductors can be made large enough to render the
precession angle $\phi = \alpha k_F \tau >2 \pi$, where $\alpha,
k_F$ are the spin-orbit coupling strength and Fermi momentum
respectively. The mathematical difficulty in obtaining the crossover
transport physics rests in the fact that one has to sum an infinite
series of diagrams which, due to the spin-orbit coupling, are not
diagonal in spin-space. In this paper we obtain the explicit
transport equations for a the series of models with spin-orbit
coupling where a special $SU(2)$ symmetry has recently been
discovered [\onlinecite{bernevig2006A}].

We first consider a two-dimensional electron gas without inversion
symmetry for which the most general form of linear spin-orbit
coupling includes both Rashba and Dresselhaus contributions:
\begin{equation}
{\cal{H}}= \frac{k^2}{2 m} + \alpha (k_y \sigma_x - k_x \sigma_y) +
\beta (k_x \sigma_x - k_y \sigma_y), \label{rashbaanddresshamilt}
\end{equation}
\noindent where $k_{x,y}$ is the electron momentum along the $[100]$
and $[010]$ directions respectively, $\alpha$, and $\beta$ are the
strengths of the Rashba, and Dresselhauss spin-orbit couplings and
$m$ is the effective electron mass. At the point $\alpha =\beta$,
which may be experimentally accessible through tuning of the Rashba
coupling via externally applied electric fields
[\onlinecite{nitta1997}], a new $SU(2)$ finite wave-vector symmetry
was theoretically discovered [\onlinecite{bernevig2006A}]. The
Dresselhauss $[110]$ model, describing quantum wells grown along the
$[110]$ direction, exhibits the above symmetry without tuning to a
particular point in the spin-orbit coupling space. At the symmetry
point, the spin relaxation time becomes infinite giving rise to a
Persistent Spin Helix. The energy bands in
Eq.[\ref{rashbaanddresshamilt}] at the $\alpha =\beta$ point have an
important {\it shifting property}: $ \epsilon_{\downarrow}(\vec{k})
= \epsilon_{\uparrow}(\vec{k} +\vec{Q})$, where $Q_+=4 m \alpha,
Q_-=0$ for the ${\cal{H_{\rm[ReD]}}}$ model and $Q_x=4 m \alpha,
Q_y=0$ for the ${\cal{H_{\rm[110]}}}$ model. The exact $SU(2)$
symmetry discovered in [\onlinecite{bernevig2006A}] is generated by
 the spin operators (written here in a transformed basis as):
\begin{eqnarray}
& S^-_{Q} = \sum_{\vec{k}} c^\dagger_{\vec{k} \downarrow} c_{\vec{k}
+ \vec{Q} \uparrow}, \;\;\; S^+_{Q} = \sum_{\vec{k}}
c^\dagger_{\vec{k}+\vec{Q}, \uparrow} c_{\vec{k} \downarrow}
\nonumber \\ & S^z_0 = \sum_{\vec{k}} c^{\dagger}_{\vec{k} \uparrow}
c_{\vec{k} \uparrow} - c^\dagger_{\vec{k} \downarrow} c_{\vec{k}
\downarrow}, \label{su2symmetry}
\end{eqnarray}
\noindent with $c_{k \uparrow, \downarrow}$  being the annihilation
operators of spin-up and down particles. These operators obey the
commutation relations for angular momentum, $[S_0^z,S^\pm_Q] = \pm 2
S_Q^\pm$  and $[S^+_Q, S^-_Q] = S_0^z$. Early spin-grating
experiments on GaAs exhibit phenomena consistent with the existence
of such a symmetry point [\onlinecite{weber2007}].

In [\onlinecite{bernevig2006A}] the spin-charge transport equations
for the Hamiltonian Eq.[\ref{rashbaanddresshamilt}] have been
obtained in the diffusive limit in which $\alpha k_F \tau << 1$.
However the regions $\alpha k_F \tau \sim 1$ and $\alpha k_F \tau >>
1$ are also experimentally accessible, and no theory is yet
available to deal with these regimes. We now present the exact spin
and charge structure factors at the exact symmetry point for any
value of the parameter $\alpha k_F \tau$.

We first obtain the spin and charge structure factors in the absence
of spin-orbit coupling, but valid in both the $\tau \rightarrow 0$
and in $\tau \rightarrow \infty$ regimes. One should think of the
structure factor obtained this way as a generalization of the
classic Lienhard formulas in the presence of disorder. We then use a
non-abelian gauge transformation introduced in
[\onlinecite{bernevig2006A}] to obtain the structure factors for the
spin-orbit coupling problem described above.

\begin{figure}
  \includegraphics[width=7cm]{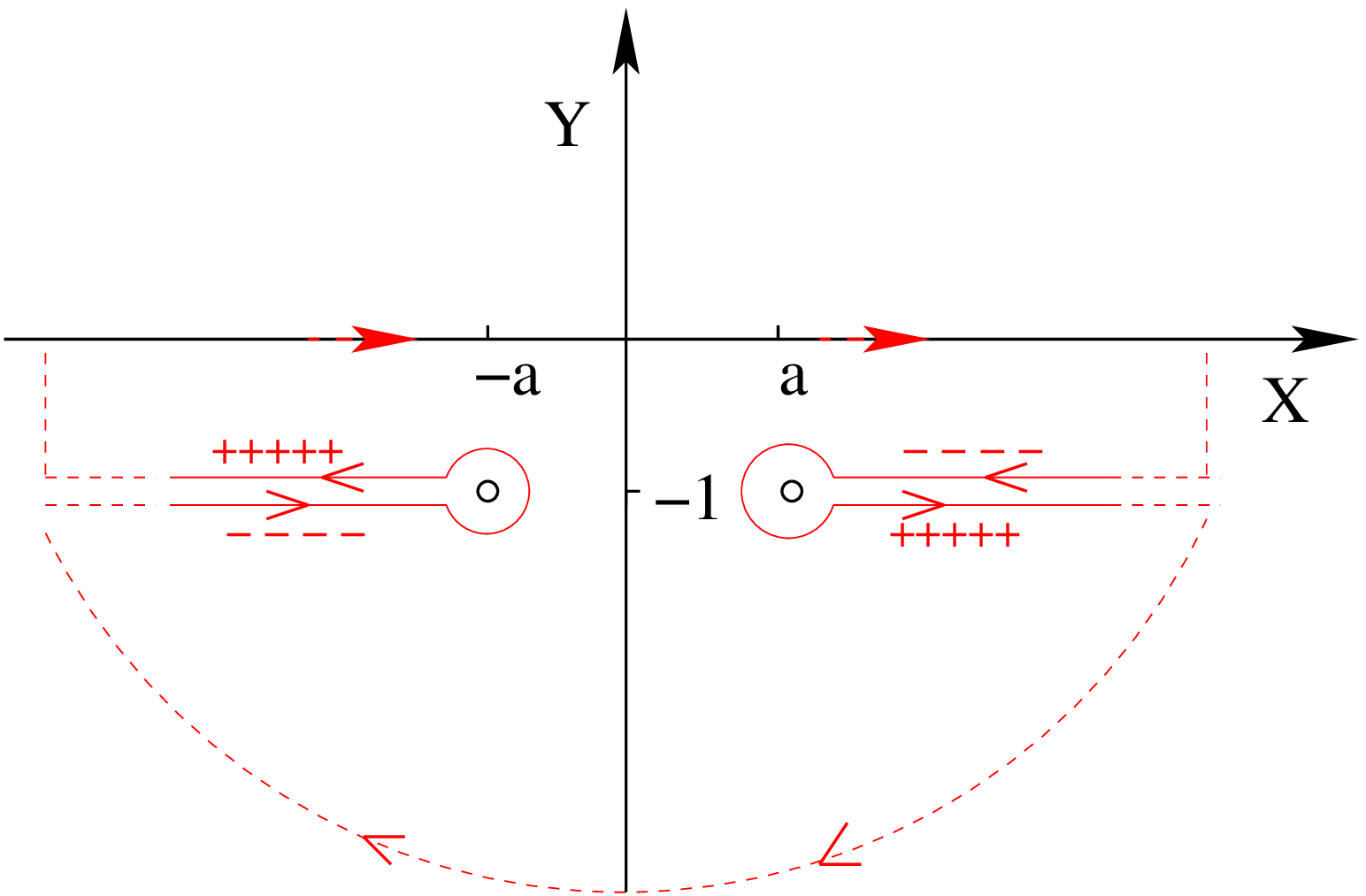}
  \caption{\label{fig_3} The sketch of the branch cut and the integral contour in the calculation of $S(t,q)$.  }
  \label{integral}
\end{figure}

We start by formulating the problem in the language of the Keyldish
formalism [\onlinecite{rammer1986,mishchenko2004}]. Assuming
isotropic scattering with momentum lifetime $\tau$, the retarded and
advanced Green's functions are:
\begin{equation}
G^{R, A}(k, \epsilon) = (\epsilon - {\cal{H}} \pm
\frac{i}{2\tau})^{-1}.
\end{equation}
\noindent We introduce a momentum, energy, and position dependent
charge-spin density which is a $2\times 2$ matrix $g(k, r,t)$.
Summing over momentum:
\begin{equation}
\rho({ r, t}) \equiv\int \frac{d^2 k}{(2 \pi)^{3} \nu} g({k,r,t}),
\end{equation}
\noindent gives the real-space spin-charge density  $\rho(r,t) =
n(r,t) + S^i(r,t) \sigma_i $, where $n(r,t)$ and $S^i(r,t)$ are the
charge and spin density and $\nu ={m}/{2 \pi} $ is the density of
states in two-dimensions. $\rho(r,t) $ and $g(k, r,t)$ satisfy a
Boltzman-type equation \onlinecite{rammer1986,mishchenko2004}:
\begin{equation}
\frac{\partial g}{\partial t} + \frac{1}{2} \left\{\frac{\partial
{\cal{H}}}{\partial k_i}, \frac{\partial g}{\partial r_i} \right\} +
i \left[{\cal{H}}, g \right] = - \frac{g}{\tau} + \frac{i}{\tau}(G^R
\rho - \rho G^A).
\end{equation}
\noindent that we now solve for a free electron gas Hamiltonian. To
obtain the spin-charge transport equations, we follow the general
sequence of technical manipulations: time-Fourier transform the
above equation; find a general solution for $g(k,r,t)$ involving
$\rho(r,t)$ and the $k$-dependent spin-orbit coupling; perform a
gradient expansion of that solution (assuming $\partial_{r} << k_F$
where $k_F$ is the Fermi wavevector) to second order; and, finally,
integrate over the momentum. The formalism is valid even through the
diffusive-ballistic boundary. For the diffusive limit, when $\tau$
is small, we need to keep only the second order term in the gradient
expansion which gives rise to the usual spin and charge propagator
$(i \omega - Dq^2)^{-1}$. As $\tau$ increases, we need to keep
higher order terms in the gradient expansion to accurately describe
the transport physics. The ballistic limit requires infinite
summation over the gradient expansion. This can be easiest seen  in
the regime of zero spin-orbit coupling, in which the sums can be
exactly performed. It is then fortuitous that our spin-orbit coupled
problem can be mapped into a free electron plus disorder problem
where we can obtain the structure factor exactly. By Fourrier
transforming in time we obtain the following recursive equation:
\begin{equation}
- i \omega \rho(r,t) = - i \int \frac{d \theta k d k}{(2 \pi)^2 m }
\Omega \sum_{n=1}^{\infty} g_n(k,r,t)
\end{equation}
\noindent were $\Omega= \omega + i/\tau$ and the $n$-th order term
reads:
\begin{equation}
g_n(k,r,t) = \partial_{r_1} ... \partial_{r_n} \left( (-
\frac{k_{i_1}}{m})...(- \frac{k_{i_n}}{m}) (\frac{i}{\Omega})^n
g_0(k,r,t) \right)
\end{equation}
\noindent where $g_0(k,r,t)$ contains a term which fixes the
momentum at the Fermi surface:
\begin{equation}
 g_0 (k,r,t)
=\frac{i}{\Omega} \frac{2 \pi}{\tau} \delta(\epsilon_F -
\frac{k^2}{2m})
\end{equation}
\noindent  Since the initial Hamiltonian and the transport equations
are rotationally invariant we can assume propagation only on $[100]$
and with the use of the identities:
\begin{equation}
\int_0^{2 \pi} d \theta (\cos(\theta))^n = \frac{( 1+ (-1)^n)
\sqrt{\pi} \Gamma(\frac{1+n}{2})}{\Gamma(1+\frac{n}{2})}
\end{equation}

\begin{equation}
 \sum_{n=1}^\infty
\frac{( 1+ (-1)^n) \sqrt{\pi}
\Gamma(\frac{1+n}{2})}{\Gamma(1+\frac{n}{2})} \frac{1}{2 \pi} a^n =
\frac{1 -\sqrt{1- a^2}}{\sqrt{1- a^2}}
\end{equation}
\noindent we can integrate over the Fermi surface angles to obtain
the structure factor pole:
\begin{equation}
S(\omega, q) = \frac{1}{ i \omega - \frac{1}{\tau} + \frac{1}{\tau}
\frac{1}{\sqrt{1 - \frac{v_F^2 q^2}{(\omega + \frac{i}{\tau})^2}}}}
\label{structurefactor}
\end{equation}
\noindent The correct interpretation of our structure factor
requires consistently picking a branch of the square-root function
in the denominator. We pick the branch cut along the positive
$x$-axis. The pole in the structure factor represents the
characteristic frequencies of the system:
\begin{equation}
\omega_{1,2} = - \frac{i}{\tau} \pm \sqrt{q^2 v_F^2 -
\frac{1}{\tau^2}}
\end{equation}
\noindent which in the diffusive and ballistic limits reduces to the
well known expressions:
\begin{eqnarray}
& \tau \rightarrow \infty \Rightarrow \omega_{1,2} \approx \pm v_F q
\nonumber \\ & \tau \rightarrow 0 \Rightarrow \omega  \approx  - i D
q^2
\end{eqnarray}
\noindent where $D = v_F^2 \tau /2$. The presence of only one
(exponentially decaying) solution in the diffusive limit follows
directly from correctly treating the branch-cut singularity in our
structure factor. It can then be seen that the exponentially
divergent solution $\omega \approx   i D q^2$ is a false pole of
Eq[\ref{structurefactor}].

Although not of immediate interest to the present paper, we also
present the structure factor for a bulk Fermi gas in the presence of
disorder. With the density of states defined as $\nu =
\frac{(2m)^{3/2} E_F^{1/2}}{ 4 \pi^2}$ the transport equation
becomes:
\begin{equation}
- i \omega \rho = - i\int\int\int \frac{d \phi \sin{\theta} d\theta
k^2 d k}{ (2 \pi)^4 \nu \tau} \Omega \sum_{n=1}^{\infty} g_n
\end{equation}
\noindent where $g_n$ and $\Omega$ are as before and $\Omega =
\omega + i/\tau$. Rotational invariance allows us to take $k_i =
k_z$ and we obtain:
\begin{widetext}
\begin{equation}
- i \Omega \rho = \frac{m k_F}{(2 \pi)^2 \nu \tau}
\sum_{n=0}^{\infty} \left(\frac{v_F q}{\Omega} \right) \int_{-1}^1
x^n d x = \frac{m k_F}{(2 \pi)^2 \nu \tau} \frac{\Omega}{v_F q}
\ln\left(\frac{1 + \frac{q}{v_F \Omega}}{1 - \frac{q}{v_F \Omega}}
\right) \rho
\end{equation}
\end{widetext}
\noindent Introducing the three-dimensional density of states at the
Fermi surface, as well as a $\delta$-function source term, the
structure factor reads:
\begin{equation}
\rho = \frac{1}{i \Omega + \frac{\Omega}{ 2 \tau v_F q}
\ln\left(\frac{1 + \frac{q}{v_F \Omega}}{1 - \frac{q}{v_F \Omega}}
\right)}
\end{equation}
\noindent To see the diffusive pole we need to carefully expand the
logarithm:
\begin{equation}
\tau \rightarrow 0: \;\;\; \rho =\frac{1}{i \omega -\frac{v_F^2
\tau}{3} q^2}
\end{equation}
\noindent Which is the right diffusive pole in $3D$. For the
ballistic pole we solve the equation (the one below is valid for any
$\tau$):
\begin{equation}
\omega = v_F q \frac{ e^{- i v_F q \tau} + e^{ i v_F q \tau} }{e^{-
i v_F q \tau} - e^{i v_F q \tau} } - \frac{i}{\tau}
\end{equation}
\noindent In the ballistic limit $\tau \rightarrow \infty$ the
exponentials in the fraction are oscillating wildly and must be
regularized. Depending on on the regularization $q \rightarrow q +
0^\pm$ the characteristic frequencies are:
\begin{equation}
\omega = \pm v_F q
\end{equation}
\noindent which are the ballistic poles.

\begin{figure}
  \includegraphics[width=6cm]{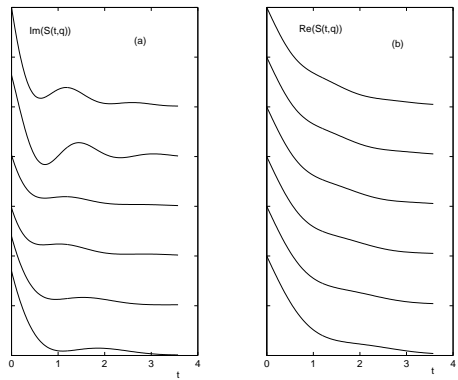}
  \caption{\label{fig_4} (a) The imaginary part   and  (b) the  real part  of $S(t,q)$. We set $\tau=1$. For both figures, from bottom to top, the curves are corresponding to $a=2.2, 2.6,
3,3.4,3.8,4.2$.  }
  \label{result}
\end{figure}

Having solved the free-Fermi gas case, we now add spin-orbit
coupling at the special $SU(2)$ symmetric point of the Persistent
Spin Helix. Following [\onlinecite{bernevig2006A}], we express the
spin-orbit coupling Hamiltonian Eq.[\ref{rashbaanddresshamilt}] in
the form of a background non-abelian gauge potential ${\cal{H_{\rm
ReD}}} = \frac{k_-^2}{2m} + \frac{1}{2m}(k_+ - 2 m \alpha
\sigma_z)^2 + const.$ where the field strength vanishes identically
for $\alpha=\beta$. Therefore, we can eliminate the vector potential
by a non-abelian gauge transformation: $\Psi_{\uparrow}(x_+,x_-)
\rightarrow \exp( i 2 m \alpha x_+) \Psi_{\uparrow}(x_+,x_-)$,
$\Psi_{\downarrow}(x_+,x_-) \rightarrow \exp(-  i 2 m \alpha x_+)
\Psi_{\downarrow}(x_+,x_-)$. Under this transformation, the
spin-orbit coupled Hamiltonian is mapped to that of the free Fermi
gas, but, while diagonal operators such as the charge $n$ and $S_z$
remain unchanged, off-diagonal operators, such as $S^-(\vec{x}) =
\psi_\downarrow^\dagger(\vec{x}) \psi_\uparrow (\vec{x})$ and
$S^+(\vec{x}) = \psi_\uparrow^\dagger(\vec{x}) \psi_\downarrow
(\vec{x})$ are transformed: $S^-(\vec{x}) \rightarrow \exp(- i
\vec{Q} \cdot \vec{r} ) S^-(\vec{x})$, $S^+(\vec{x}) \rightarrow
\exp( i \vec{Q} \cdot \vec{r}) S^+(\vec{x})$. Here $\vec{Q}$ is the
shifting wavevector of the spin-orbit coupled Hamiltonian. Since in
the gauge transformed basis, all three components of the spin and
charge have the structure factor derived above, in the original
(experimentally measurable) basis, the $S_x$ and $S_y$ have the
following form:
\begin{equation}
S^\pm(\omega, \vec{q}) = \frac{1}{ i \omega - \frac{1}{\tau} +
\frac{1}{\tau} \frac{1}{\sqrt{1 - \frac{v_F^2 (\vec{q} \pm
\vec{Q})^2}{(\omega + \frac{i}{\tau})^2}}}}
\end{equation}
\noindent The above result represents the exact form factor for a
spin-orbit coupled system valid everywhere from the diffusive to
ballistic regimes. The Persistent Spin Helix is clearly maintained
for any values of $\tau, \alpha, v_f$ since $S(\omega, \vec{Q}) =
1/i \omega$ which renders the spin life-time infinite.

The transient grating experiments [\onlinecite{weber2007,gedik2003}]
measure the $\omega$ Fourrier transform of $S(\omega,q)$, i.e.
$S(t,q) = \frac{1}{2\pi}\int dt e^{-i\omega t} S(\omega,q)$.
$S(\omega, q)$ is analytic in the upper half complex plane. Thus,
$S(t,q)$ is zero for $t<0$. For $t>0$, by selecting the integral
contour as shown in fig.(\ref{integral}), we obtain its real part
and imaginary part as follows:
\begin{eqnarray}
&
 &\frac{Im(S(t,q))}{e^{-\frac{t}{\tau}}}=\frac{a}{1+a^2}+P\int_{a}^{\infty}\frac{2}{\pi}\frac{\sqrt{x^2-a^2}cos
(\frac{xt}{\tau})}{x(x^2-1-a^2)} \nonumber
 \\
& & \frac{Re (S(t, q))}{e^{-\frac{t}{\tau}}}=-
\frac{a^2+cos(\sqrt{1+a^2}\frac{t}{\tau})}{1+a^2}
\end{eqnarray}
\noindent where $a=v_F|\vec{q}\pm \vec{Q}|\tau$ and $P$ indicates
the principal value of the integral.

\begin{figure}
  \includegraphics[width=6cm]{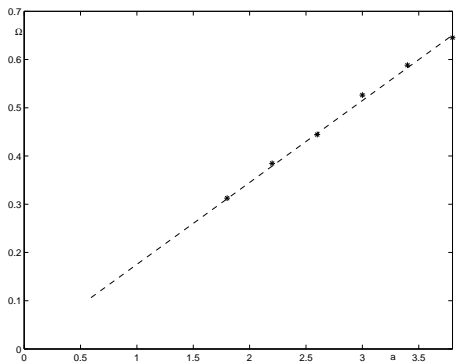}
  \caption{\label{fig_5}  The oscillation frequency $\Omega$ in the imaginary part of $S(t,q)$ as a function of $a=\nu_F|\vec q\pm\vec Q|\tau$. }
  \label{linear}
\end{figure}

In Fig.(\ref{result}), we plot the real and imaginary part of
$S(t,q)$ for different values of $a$. In the figure, we set $\tau=1$
and from bottom to top, the curves are corresponding to $a=2.2, 2.6,
3,3.4,3.8,4.2$. Although the real part is clearly an oscillating
function of $t$ with an oscillation frequency,
$\frac{\sqrt{1+a^2}}{\tau}$, the oscillation is not easily seen in
the figure. However, the imaginary part has a much larger
oscillation amplitude than the real part and the oscillation becomes
clear as increasing $a$, reflecting the ballistic nature of the
sample. The oscillation frequency $\Omega$ in the imaginary part is
linearly dependent on $a$ as shown in Fig.(\ref{linear}).

In this paper we have obtained the exact transport equations valid
in the diffusive, ballistic, and crossover regimes of a special type
of spin-orbit coupled system which enjoys an $SU(2)$ gauge symmetry.
We obtained the exact form of the structure factors, and found the
dependence of the spin-density as would be observed in a
transient-grating experiment. It would be interesting to work out
the transport equations in the diffusive-ballistic regime in
perturbation theory away from the Persistent Spin Helix.

B.A.B. wishes to acknowledge the hospitality of the Kavli Institute
for Theoretical Physics at University of California at Santa
Barbara, where part of this work was performed. BAB acknowledges
fruitful discussions with Joe Orenstein, C.P. Weber, Jake Koralek
and Shoucheng Zhang. This work is supported by the Princeton Center
for Theoretical Physics and   by the National Science Foundation
under grant number: PHY-0603759.

\bibliography{diffusiveballisticbib}

\end{document}